\documentclass[12pt,preprint]{aastex}

\usepackage{apjfonts}
\usepackage{emulateapj5}
\usepackage{amsmath}
\usepackage{epsfig}
\usepackage{amssymb}

\newcommand{\ptl}{\partial}

\renewenvironment{figure}{\begin{figure*} }{\end{figure*}}

\newcommand{\bu}{{\bf u}}

\newcommand{\br}{{\bf r}}
\newcommand{\grad}{{\mathbf \nabla}}
\renewcommand{\div}{{\mathbf \nabla} \cdot}

\newcommand{\bO}{\mbox{\boldmath $\Omega$}}

\begin{document}

\title{On the penetration of meridional circulation below the solar convection zone}

\author{P. Garaud \& N. H. Brummell} 

\affil{Department of Applied Mathematics and Statistics, Baskin School of Engineering, University of California Santa Cruz, 1156 High Street, CA 95064 
Santa Cruz, USA}

\maketitle

\begin{abstract}
Meridional flows with velocities of a few meters per second are observed 
in the uppermost regions of the solar convection zone. The amplitude 
and pattern of the flows deeper in the solar interior, in particular near 
the top of 
the radiative region, are of crucial importance to a wide range of 
solar magnetohydrodynamical processes. In this paper, we provide a systematic
study of the penetration of large-scale meridional flows from the convection 
zone into the 
radiative zone. In particular, we study the effects of the assumed
boundary conditions applied at the convective-radiative interface on the 
deeper flows. Using simplified analytical models in conjunction with more 
complete numerical methods, 
we show that penetration of the convectively-driven meridional flows into 
the deeper
interior is not necessarily limited to a shallow Ekman depth but can 
penetrate much deeper, 
depending on how the convective-radiative interface flows are modeled. 
\end{abstract}


\section{Introduction}

Meridional flows in the solar interior have recently become the focus 
of observational and theoretical attention. Poleward sub-surface 
flows with amplitudes of the order of a few tens of meters per second have
been detected with reliable accuracy down to
about 0.85$r_\odot$ (Giles {\it et al.} 1997).  
A globally equatorward
return flow must exist deeper in the interior to guarantee mass conservation, 
but its amplitude and structure can only be conjectured currently. 
This paper addresses the question of 
how deeply these meridional flows penetrate into the radiative zone.
An understanding of these return meridional flows is of fundamental importance
since their nature plays a crucial role in many current theories for 
the internal magneto-hydrodynamics of the Sun.

Firstly, meridional circulations have
been argued to play a central role in the operation of the global 
solar dynamo (see the review by 
Charbonneau, 2005). In these models, 
the predicted spatio-temporal behavior of the solar cycle depends
sensitively on the assumed circulation pattern and speed.
The chosen position for the return flow coupled with mass conservation
sets the velocity of the equatorward flow near the base of the convection 
zone and therefore controls the activity cycle period.
Similarly, the depth of penetration of the meridional 
flows into the radiative 
region determines where the toroidal magnetic field is generated, 
and therefore also influences the cycle period and the field amplitudes.

Secondly, meridional flows advect angular momentum, and therefore play
a key role in the global dynamical balance of the solar interior. 
For example, helioseismology has revealed the existence of a strong radial 
shear layer, now known as the solar tachocline (Brown {\it et al.} 1989;
Spiegel \& Zahn, 1992; Hughes, Rosner \& Weiss, 2007),
located precisely at the interface between the radiative and convective 
regions. Quantitative models of the tachocline have revealed a sensitive 
dependence of the interior angular velocity profile on the derived or 
assumed interfacial flows (Spiegel \& Zahn, 1992; Gough \& McIntyre, 1998; 
Rempel, 2005; Garaud, 2007).

Finally, meridional flows also transport various chemical species within the
solar interior, with directly observable consequences. 
Near the base of the convection zone, mixing by 
large-scale flows can prevent the 
gravitational settling of helium with respect to hydrogen, leaving a 
noticeable signature in the helioseismic sound-speed data (Elliott \& Gough ,1999). 
Additional mixing below the convection zone in the main-sequence phase is also 
required by the observed surface abundances of light elements such as 
lithium and beryllium, with plausibly the same origin.

Flows with amplitudes on the order of tens of meters per second 
are required to balance angular momentum transport by the turbulent stresses 
throughout the convection zone
(Miesch {\it et al}. 2000). Since there exist no physical barrier 
between the convective and
radiative regions, these convectively driven flows may continue 
their downward progress somewhat beyond the driving region into the 
radiative zone, thereby ``penetrating'' or ``overshooting'' into the interior
while retaining potentially significant velocities. 
How far flows extend beyond their driving region 
is an essential question. 

The problem
has recently been addressed by Gilman \& Miesch (2004) 
(GM04 hereafter) and by McIntyre (2007). Using a steady-state formalism, 
GM04 discuss the depth of penetration of an existing 
{\it latitudinal} flow into the radiative interior. They argue that the source flow 
amplitude is rapidly damped in the radiative interior within a shallow Ekman depth.
This depth ranges from a fraction of a kilometer, using microscopic values of 
the viscosity, to a few tens of kilometers, using a turbulent value of the 
viscosity. Gilman and Miesch reach a strong conclusion, namely that 
{\it ``the physics of the solar tachocline and neighboring regions does not 
allow penetration of meridional circulation originating in the solar 
convection zone below the overshoot layer''}.  This could have 
dramatic consequences for the magneto-hydrodynamics of the solar 
radiative zone. 

However, steady-state solutions are sensitively dependent on boundary 
conditions.  GM04 solve the problem for the radiative interior dynamics 
where the forcing is by purely latitudinal flows at the upper 
convective-radiative
interface.  They do not consider interfacial forcing from flows 
generated by the differential 
rotation nor by direct radial pumping into the radiative zone 
(only radial flows
generated for mass conservation in response to their latitudinal forcing).  
It is reasonable to address the question as to whether their strong 
conclusion remains applicable under more general circumstances.

In this paper, we therefore extend the work of GM04 
to allow for greater generality in the source forcing flow, allowing
the possibility of azimuthal and radial flows in addition to the
latitudinal flows.  We systematically examine the consequences of using 
these various sets of boundary conditions to mimic the convective-radiative
interface. We find that the meridional flows can penetrate to 
significantly different depths, depending upon the choice of 
boundary conditions.  The
GM04 solution can be recovered in special cases, but is typically 
overpowered by other solutions when alternative assumptions are made about 
the nature of the flows driven within the convection zone.

In what follows, we adopt a three-step approach to study the 
penetration of meridional flows into the 
solar radiative zone. Given the added difficulties inherent to spherical
coordinate systems, we first examine the problem in Cartesian geometry. 
In \S \ref{sec:cart}, we 
study analytically the complete set of steady, linear, pseudo-axisymmetric Boussinesq equations 
and explore a wide range of boundary conditions that may possibly mimic 
the effects of the convection zone on the radiative zone. We systematically
discuss the solutions obtained, which are linear 
combinations of two fundamental modes of behavior.  One of these modes is  
a solution which varies rapidly on a typical shallow Ekman scale 
as found by GM04, 
the other one is a more slowly varying solution which can span 
the entire interior. In order to gain better insight into the physics
of the system and in particular the new solutions, in \S \ref{sec:toy} we
consider a simplified Cartesian 
Boussinesq model in which the solutions related to the Ekman layers 
are artificially suppressed by neglecting the viscous terms in the radial and 
latitudinal components of the momentum equation. 
Finally, in \S \ref{sec:num}, we relax the Cartesian constraint and present 
numerical results for all of the
various boundary conditions in a steady, linear, anelastic, spherical but 
axisymmetric
simulation of the solar radiative zone. We compare these numerical results in
spherical geometry 
with the analytical predictions from the Cartesian models, both in terms
of the scale of variation of the solutions, and in terms of their predicted 
flow velocities. While the spherical geometry as well as the non-uniform 
background state necessarily add to the complexity of the problem, we find 
that the analytical scalings extracted from the Cartesian geometry models
agree very well with the full numerical solutions. When more 
complex boundary conditions are taken into account, limits on the 
penetration of meridional flows into the radiative zone are much less 
stringent than previously claimed. The exact flow velocities and depth of 
penetration achieved, however, depend sensitively on the actual 
boundary conditions selected. These results and conclusions are discussed in detail  
in \S\ref{sec:ccl}.

\section{A Cartesian model}
\label{sec:cart}

In all that follows we consider a stably stratified radiative zone
located beneath a turbulent convection zone, a typical situation
encountered in all solar-type stars. Within the convection zone, we
assume that turbulent stresses drive large-scale flows in the
azimuthal direction (i.e. a large-scale differential rotation) as well as in
the meridional direction. The amplitude and spatial variation of the
flows just above the convective-radiative interface is assumed to be
known. We then pose and solve the following question: {\it What is the
resulting flow pattern and velocities in the underlying radiative zone?}

\subsection{Model setup}

The spherical geometry of the solar radiative zone as well as the 
non-uniform background state (in terms of temperature and density, viscosity 
and thermal conductivity for instance) both preclude any attempt at solving 
the problem analytically. We postpone to \S\ref{sec:num} the presentation
and discussion of the complete numerical solution of the problem, and 
first consider a much simplified ``radiative zone'' with rectangular geometry 
in Cartesian coordinates $(x,y,z)$, and a uniform background temperature 
gradient, viscosity and thermal conductivity. In this coordinate system,
the $x$-direction can be thought of as the azimuthal direction  with 
$x \in [0,2\pi R]$, the $y$-direction is aligned with the latitudinal 
direction and is limited to $y \in [0,\pi R]$ and finally the $z$-direction is 
the radial direction with $z \le R$. The poles are represented by $y=0$ and 
$y = \pi R$ while the equator is at $y = \pi R/2$. The dimensional 
constant $R$ represents the base of the convection 
zone, and $z=0$ the interior of the Sun. The system rotates with angular 
velocity $\bO = (0,0,\Omega)$, and gravity is assumed to be aligned with 
the rotation axis. 

\subsection{Model equations and general solution}

The equations governing dynamical and thermal perturbations to a 
stably stratified background assumed at rest are the mass, 
momentum and thermal energy conservation equations. Using the Boussinesq 
approximation and assuming ``axial'' symmetry 
(i.e. $\partial /\partial x = 0$), we first linearize these equations 
in the thermal perturbations and flow velocities, then project them onto 
the Cartesian coordinate system as
\begin{eqnarray}
&&\frac{\ptl v}{\ptl y} + \frac{\ptl w}{\ptl z} = 0 \mbox{   ,   } \nonumber \\
&&-2 \Omega v = \nu \left( \frac{\ptl^2 u}{\ptl y^2}+ \frac{\ptl^2 u}{\ptl z^2} \right) \mbox{   ,   }  \nonumber \\
&&2 \Omega u = - \frac{\partial p}{\partial y} + \nu \left( \frac{\ptl^2 v}{\ptl y^2}+ \frac{\ptl^2 v}{\ptl z^2} \right) \mbox{   ,   }  \nonumber \\
&&0 = - \frac{\partial p}{\partial z} + g \alpha \theta  \mbox{   ,   }  \nonumber \\
&&\beta w = \kappa \left( \frac{\ptl^2 \theta}{\ptl y^2} + \frac{\ptl^2 \theta}{\ptl z^2} \right) \mbox{   ,   } 
\label{eq:maineqs}
\end{eqnarray}
where $\bu = (u,v,w)$ is the flow velocity, $\theta$ is the
temperature perturbation, $\nu$ and $\kappa$ are the viscosity and
thermal diffusivity, $\alpha = 1/\overline{T}$ where $\overline{T}$ is 
the background temperature and finally, $g \alpha \beta = N^2$ is the 
background buoyancy frequency. Note that the viscous diffusion term in the 
radial component of the momentum equation has been neglected in accordance
with hydrostatic equilibrium; this does not affect in any way the conclusions
of this paper.

While GM04 and McIntyre (2007) neglected the $\partial^2/\partial y^2$ terms 
in the viscous terms, we consider here for completeness the full Laplacian. 
The additional terms are 
found to be necessary in the light of the fact that some of the boundary 
layers 
in the system are large compared with the vertical size of the domain. 
Neglecting these terms is not physically justified, and mathematically
transforms any slowly varying, standard exponential solutions 
into the rather unphysical secular linear solutions described by McIntyre 
(2007) (and neglected by GM04).

The hydrostatic equilibrium equation can be combined with the latitudinal 
component of the momentum equation to yield a generalized thermal-wind 
equation, 
\begin{equation}
2 \Omega \frac{\ptl u}{\ptl z} + \alpha g \frac{\ptl \theta}{\ptl y}  
= \nu \frac{\partial}{\partial z} \left( \frac{\ptl^2 v}{\ptl y^2} + 
\frac{\ptl^2 v}{\ptl z^2} \right) \mbox{   .   } 
\end{equation}

Seeking solutions with latitudinal dependence as $\sin(2n y /R)$ or 
$\cos(2n y /R)$, and exponential vertical dependence as $\exp(kz)$, 
we obtain the characteristic equation 
\begin{equation}
\left(k^2 - \frac{4n^2}{R^2} \right) \left[  k^2\left(k^2 - \frac{4n^2}{R^2}
\right)^2 + \frac{k^2}{d_{\rm E}^4} -  \frac{4n^2}{R^2 d_{\rm BD}^4}\right]  
= 0 \mbox{   ,   } 
\label{eq:caract1}
\end{equation}
where we have introduced two standard characteristic lengthscales
\begin{equation}
d_{\rm E} = \left(\frac{\nu}{2\Omega}\right)^{1/2} \mbox{   and   } 
d_{\rm BD} =\left( \frac{\nu\kappa}{N^2} \right)^{1/4}  \mbox{   ,   } 
\end{equation}
the first one being the standard Ekman depth and the second representing a 
buoyancy-diffusion layer (GM04, Barcildon \& Pedlosky 1967).

It is possible to show (with some algebra) that in the limit where 
\begin{equation}
\frac{R^4 d_{\rm BD}^8}{16 n^4 d_{\rm E}^{12}}  \gg 1  \mbox{   ,   } 
\end{equation}
(which is always true below the base of the convection zone) then the 
eight solutions to equation (\ref{eq:caract1}) ($\pm k_1$, $\pm k_2$, 
$\pm k_3$ and $\pm k_4$) can be approximated by 
\begin{eqnarray}
&&k_1 = \frac{2n}{R}  \mbox{   ,   }  \nonumber \\
&&k_{2} \simeq \frac{d_{\rm E}^2}{d_{\rm BD}^2} \frac{2n}{R}  \mbox{   ,   }  
\nonumber \\
&&k_{3,4} \simeq \frac{\sqrt{2}}{2} (1\pm i) d_{\rm E}^{-1}  \mbox{   ,   } 
\label{eq:kis}
\end{eqnarray}
thus yielding four slowly varying exponential solutions (related to $k_1$ 
and $k_2$) in addition to four rapidly varying, oscillatory and exponential 
solutions related to $k_3$ and $k_4$ which describe the Ekman layers of the 
system. The Ekman solutions were found and described by GM04. On the other 
hand, the $\pm k_1$ and $\pm k_2$ solutions differ from the (two) real 
solutions found by GM04, a discrepancy which can easily be traced back 
to the omitted $\partial^2/\partial y^2$ in their viscous diffusion terms. 

It is interesting to note that 
\begin{equation}
k_2 = \sqrt{Pr Bu} \frac{n}{D} = \sqrt{Pr Bu} \frac{R}{2D} k_1 \mbox{   ,   } 
\label{eq:k2}
\end{equation}
where $D$ is the local density scaleheight, $Pr=\nu/\kappa$ is the Prandtl 
number and where the Burger number $Bu$ is defined as
\begin{equation}
Bu = \left( \frac{N D}{R \Omega} \right)^2 \mbox{   .   } 
\end{equation}
 The Burger number in the solar radiative zone is estimated to be about 
$Bu \simeq 2.5\times 10^3$ using $\Omega = 2.7\times 10^{-6}$s$^{-1}$, 
$N = 8 \times 10^{-4}$s$^{-1}$, $R = 5 \times 10^{10}$cm and 
$D = 0.17 R = 8.6 \times 10^{9}$cm (see Gough, 2007). In laminar regions 
of the solar radiative zone, the microscopic Prandtl number is of order 
$Pr = 2 \times 10^{-6}$, so that
\begin{equation}
k_2^{-1} \simeq 5 k_1^{-1} \mbox{   ,   } 
\label{eq:myd}
\end{equation}
which is clearly of the order of $R$ itself for large-scale forcing 
(small $n$).

\subsection{General solutions}

Consider for instance the solution of (\ref{eq:maineqs}) corresponding to 
$\cos(2ny/R)$ and $\sin(2ny/R)$ variations in the azimuthal velocity:
\begin{eqnarray}
u(x,y,z) &=& \left( a_1e^{k_1z}  + a_2 e^{-k_1z} + a_3 e^{k_2z} 
+ a_4 e^{-k_2z} \right. \\ 
 &+& \left. a_5 e^{k_3z} + a_6 e^{-k_3z} + a_7 e^{k_4z} 
+ a_8 e^{-k_4z} \right) \cos\left(\frac{2 ny }{R} \right) \nonumber \\
 &+& \left( b_1e^{k_1z}  + b_2 e^{-k_1z} + b_3 e^{k_2z} 
+ b_4 e^{-k_2z} \right. \nonumber  \\ 
 &+& \left. b_5 e^{k_3z} + b_6 e^{-k_3z} + b_7 e^{k_4z} 
+ b_8 e^{-k_4z} \right) \sin\left(\frac{2 ny }{R} \right)  \mbox{   .   }  
\nonumber\end{eqnarray}
This corresponds to 
\begin{eqnarray}
v(x,y,z) &=&  d_{\rm E}^2 \left[ \left( k_1^2 - k_2^2 \right) 
\left(a_3 e^{k_2z} + a_4  e^{-k_2z} \right) \right.  \nonumber \\ 
 &+&  \left( k_1^2 - k_3^2 \right) \left( a_5 e^{k_3z} 
+ a_6 e^{-k_3z} \right) \nonumber \\ &+& \left. 
\left( k_1^2-k_4^2\right) \left(  a_7 e^{k_4z} 
+ a_8  e^{-k_4z} \right) \right] \cos\left( \frac{2n y }{R} \right) 
\nonumber \\ 
 &+&  d_{\rm E}^2 \left[ \left( k_1^2 - k_2^2 \right) \left(b_3 e^{k_2z} 
+ b_4  e^{-k_2z} \right) \right.  \nonumber \\ 
 &+&  \left( k_1^2 - k_3^2 \right) \left( b_5 e^{k_3z} + b_6 e^{-k_3z} 
\right) \nonumber \\ &+& \left. \left( k_1^2-k_4^2\right) \left(  b_7 e^{k_4z} 
+ b_8  e^{-k_4z} \right) \right] \sin\left( \frac{2n y }{R} \right)  
\mbox{   ,   } 
\end{eqnarray}
\begin{eqnarray}
w(x,y,z) &=& d_{\rm E}^2 \left[ \frac{k_1}{k_2}\left( k_1^2 - k_2^2 \right) 
\left(a_3 e^{k_2z} - a_4  e^{-k_2z} \right) \right. \nonumber  \\ 
 &+&  \frac{k_1}{k_3} \left( k_1^2 - k_3^2 \right) \left( a_5 e^{k_3z} 
- a_6 e^{-k_3z} \right) \nonumber \\ &+& \left. \frac{k_1}{k_4} \left( k_1^2 - 
k_4^2\right) \left(  a_7 e^{k_4z} - a_8  e^{-k_4z} \right) \right] 
\sin\left( \frac{2n y }{R} \right)  \nonumber \\ 
&-& d_{\rm E}^2 \left[ \frac{k_1}{k_2}\left( k_1^2 - k_2^2 \right) 
\left(b_3 e^{k_2z} - b_4  e^{-k_2z} \right) \right. \nonumber  \\ 
 &+&  \frac{k_1}{k_3} \left( k_1^2 - k_3^2 \right) \left( b_5 e^{k_3z} 
- b_6 e^{-k_3z} \right) \nonumber \\ &+& \left. \frac{k_1}{k_4} 
\left( k_1^2 - 
k_4^2\right) \left(  b_7 e^{k_4z} - b_8  e^{-k_4z} \right) \right] 
\cos\left( \frac{2n y }{R} \right)  \mbox{   ,   } 
\end{eqnarray}
and finally 
\begin{eqnarray}
\theta(x,y,z) &=& - \frac{2\Omega}{g\alpha }\left[ \left( a_1 e^{k_1z} -a_2 e^{-k_1z} \right) + \frac{k_2}{k_1} \left(a_3 e^{k_2z} - a_4  e^{-k_2z} 
\right)  \right. \nonumber \\
&+& \left. \frac{k_2^2}{k_1k_3} \left(a_5 e^{k_3z} - a_6  e^{-k_3z} \right) +  \frac{k_2^2}{k_1k_4} \left(a_7 e^{k_4z} - a_8  e^{-k_4z} \right) 
\right] \nonumber \\
& \cdot &  \sin\left( \frac{2ny}{R} \right)  \nonumber \\
 &+& \frac{2\Omega}{g\alpha }\left[ \left( b_1 e^{k_1z} -b_2 e^{-k_1z} \right) + \frac{k_2}{k_1} \left(b_3 e^{k_2z} - b_4  e^{-k_2z} \right)  \right. 
\nonumber \\
&+& \left. \frac{k_2^2}{k_1k_3} \left(b_5 e^{k_3z} - b_6  e^{-k_3z} \right) +  \frac{k_2^2}{k_1k_4} \left(b_7 e^{k_4z} - b_8  e^{-k_4z} \right) 
\right] \nonumber \\
& \cdot &  \cos\left( \frac{2ny}{R} \right)  \mbox{   ,   } 
\end{eqnarray}
where one must bear in mind that $k_1$ and $k_2$ depend on the wavenumber $n/R$ of the forcing function.

\subsection{Boundary conditions}
\label{subsec:bcs}

To find the amplitude of each exponential term, one must apply the 
boundary conditions to the 
general solutions. A quick look at the system 
shows that eight boundary conditions are required, which arise from 
conditions on $u$, $v$, $w$ and $\theta$ at both the top and bottom 
of the domain. 

We choose boundary conditions at $z = R$ to represent the action of
the convection zone on the underlying stably stratified and laminar radiative 
region. We require the continuity of the radiative interior 
solution with the complete vector of velocities at the base of the convection 
zone, so that
\begin{equation}
\bu(x,y,R) = \bu_{\rm cz}(y) = (u_{\rm cz}(y),v_{\rm cz}(y),w_{\rm cz}(y))
\end{equation}
(where the meridional and azimuthal flows in the convection zone are 
assumed to be axisymmetric). A simple reasonable prescription for 
the flow velocities at the interface might be, for instance,
\begin{eqnarray}
&&u_{\rm cz}(y) = - U_0 \cos\left(\frac{2y}{R}\right) \mbox{   ,   }  \nonumber \\
&&v_{\rm cz}(y) = V_0 \sin\left(\frac{2y}{R}\right)  \mbox{   ,   } \nonumber \\
&&w_{\rm cz}(y) = -W_0 \cos\left(\frac{2y}{R}\right)  \mbox{   .   } 
\end{eqnarray}
The azimuthal forcing term $u_{\rm cz}(y)$ represents 
a solar-like differential rotation (with slower-than-average 
rotation near the poles and faster-than-average rotation near the equator 
if $U_0 > 0$). The latitudinal and radial forcing terms 
$v_{\rm cz}(y)$ and $w_{\rm cz}(y)$ represent 
a single-cell flow in each hemisphere with independent vertical 
and latitudinal flow velocities. When $V_0 > 0$ the flow is equatorward 
at the boundary, while $W_0 > 0$ guarantees inflow in the high latitudes 
and outflow in the low latitudes. Note that since the problem 
studied is linear, and since any of the three velocity components of 
$(u_{\rm cz}(y),v_{\rm cz}(y),w_{\rm cz}(y))$ can be written as a Fourier 
series in $\cos(2ny/R)$ and $\sin(2ny/R)$, it is possible to find the 
general solution of (\ref{eq:maineqs}) for any set of imposed velocities. The 
profiles chosen here contain only one Fourier component for simplicity.

Near the bottom boundary, we would in principle like to choose boundary 
conditions that have as little effect on the solution as possible. 
In the real Sun they would instead be replaced by regularity 
conditions at the origin. However, fitting eight boundary conditions to 
the general solutions yields an 8$\times$8 linear system which is difficult to
study analytically (even with the help of Maple or equivalent software).
Thus in this section we restrict our study to the case where
the bottom boundary is located at $z \rightarrow -\infty$. This immediately
implies that all of the even $\{a_i\}$ and $\{b_i\}$ coefficients must be null.
In \S\ref{sec:toy} we revisit this simplification, and suggest 
an easy way of deducing the solutions in a geometry where the bottom 
boundary is at $z=0$ from those in a semi-infinite 
domain. In \S\ref{sec:num} we verify our hypothesis against numerical 
results.

In what follows (\S\ref{subsec:case1} and \S\ref{subsec:case2}) 
we describe two possibilities 
for the thermal boundary conditions at the convective-radiative interface.

\subsection{Thermal boundary conditions of type 1: ``perfectly'' 
conducting convection zone}
\label{subsec:case1}

The local heat flux through the boundary associated with the perturbations 
is the sum of the conducted heat flux $- k \partial \theta /\partial z$ 
and the advected heat flux $\overline{\rho} \overline{h} w$, where 
$\overline{\rho}$ is the background density, $\overline{h}$ is the background 
enthalpy, and $k = \overline{\rho} c_{\rm p} \kappa$ is the 
thermal conductivity (here $c_{\rm p}$ is the specific heat at constant 
pressure). 

In a steady-state this local heat flux must be equal to zero, so that when 
the convection zone is perfectly conducting ($k \rightarrow \infty$) 
the thermal boundary condition reduces to $\partial \theta / \partial z = 0$. 

Using this final boundary condition together with the ones described in 
\S\ref{subsec:bcs} we find that 
the remaining odd coefficients $\{a_i\}_{i=1,3,5,7}$ satisfy the linear system 
$M A = C $ with 
\begin{equation}
M = \begin{pmatrix}
1 & 1 & 1 & 1\\
0 & k_1^2-k_2^2 & k_1^2-k_3^2 & k_1^2 - k_4^2 \\
0 & \frac{k_1^2-k_2^2}{k_2} & \frac{k_1^2-k_3^2}{k_3} & \frac{k_1^2 - k_4^2}{k_4}  \\
k_1^2 & k_2^2 & k_2^2 & k_2^2
\end{pmatrix} 
\end{equation}
and 
\begin{equation}
A = \begin{pmatrix} a_1e^{k_1R} \\ a_3e^{k_2R} \\  a_5e^{k_3R} \\ a_7e^{k_4R} 
\end{pmatrix} \mbox{ and   } C = \begin{pmatrix} -U_0 \\ 0 \\ 0 \\ 0 \end{pmatrix} \mbox{   .   } 
\end{equation}
Similarly, the coefficients $\{b_i\}_{i=1,3,5,7}$ satisfy the system  $MB = D$ with  
\begin{equation}
B = \begin{pmatrix} b_1e^{k_1R} \\ b_3e^{k_2R} \\  b_5e^{k_3R} \\ b_7e^{k_4R} 
\end{pmatrix} \mbox{ and   } D = \begin{pmatrix} 0 \\ V_0d_{\rm E}^{-2} \\ W_0 d_{\rm E}^{-2} k_1^{-1} \\ 0 \end{pmatrix} \mbox{   .   } 
\end{equation}

In the limit where $k_2 \ll k_1 \ll |k_3|, |k_4|$ the solutions can be simplified to yield 
\begin{eqnarray}
&&a_1 \simeq U_0 \frac{k_2^2}{k_1^2} e^{-k_1R}  \mbox{   ,   }  \nonumber \\
&&a_3 \simeq  - U_0 e^{-k_2R}  \mbox{   ,   }  \nonumber \\
&&a_5 \simeq U_0 \frac{k_1^2}{k_2}\frac{k_4}{k_3(k_3-k_4)} e^{-k_3R}  \mbox{   ,   } \nonumber \\
&&a_7 \simeq U_0 \frac{k_1^2}{k_2} \frac{k_3}{k_4(k_4-k_3)} e^{-k_4R}  \mbox{   ,   } 
\end{eqnarray}
and
\begin{eqnarray}
&&b_1 \simeq 0  \mbox{   ,   }  \nonumber \\
&&b_3 \simeq  \left[ -\frac{V_0}{d_{\rm E}^2} \frac{1}{k_3k_4}  
+ \frac{W_0}{ d_{\rm E}^2}  \frac{(k_3+k_4)}{k_1k_3k_4} \right] e^{-k_2R}
 \mbox{   ,   }  \nonumber \\
&&b_5 \simeq \left[- \frac{V_0}{d_{\rm E}^2} \frac{1}{k_3(k_3-k_4)} 
+ \frac{W_0}{d_{\rm E}^2 k_1} \frac{k_4}{k_3(k_3-k_4)} \right] e^{-k_3R} 
 \mbox{   ,   }  \nonumber \\
&&b_7 \simeq \left[ -\frac{V_0}{d_{\rm E}^2} \frac{1}{k_4(k_4-k_3)}  
+ \frac{W_0}{d_{\rm E}^2k_1} \frac{k_3}{k_4(k_4-k_3)} \right] e^{-k_4R}  \mbox{   .   } 
\end{eqnarray}

We can now finally evaluate, for instance, the latitudinal flow velocity 
within this rectangular radiative zone:
\begin{eqnarray}
v(x,y,z) &\simeq& \left[ - d_{\rm E}^2 k_1^2 U_0 e^{k_2(z-R)} 
- d_{\rm E}^2 k_1^2 \frac{U_0}{k_2} \frac{k_4k_3}{k_3-k_4} e^{k_3(z-R)}  
\right. \\ &-& \left. d_{\rm E}^2 k_1^2 \frac{U_0}{k_2} 
\frac{k_3k_4}{k_4-k_3}  e^{k_4(z-R)}  \right] \cos\left( \frac{2 y }{R} \right)
 \nonumber \\ 
&+& \left[ \left( - V_0 \frac{k_1^2}{k_3k_4} + W_0\frac{k_1 (k_3+k_4)}{k_3k_4}
 \right) e^{k_2(z-R)} \right.  \nonumber \\ &+&  \left( V_0 \frac{k_3}{k_3-k_4} 
- \frac{W_0}{k_1} \frac{k_4k_3}{k_3-k_4} \right)  e^{k_3(z-R)} \nonumber \\ 
&+& \left.  \left( V_0 \frac{k_4}{k_4-k_3} - \frac{W_0}{k_1} 
\frac{k_3k_4}{k_4-k_3} \right) e^{k_4(z-R)}  \right] 
\sin\left( \frac{2 y }{R} \right)  \mbox{   .   }  \nonumber 
\label{eq:vcase1}
\end{eqnarray}
As expected from the linearity of the governing system equations 
(\ref{eq:maineqs}), 
the meridional flow velocity in the radiative interior is simply the sum of 
the three contributions arising from azimuthal forcing only (with terms 
proportional to $U_0$), latitudinal forcing only 
(with terms proportional to $V_0$) and radial forcing only 
(with terms proportional to $W_0$). In addition, each of these three
contributions is the sum of three terms, one with exponential dependence
in $e^{k_2(z-R)}$ which corresponds to a very slowly varying function of depth,
and two complex conjugate terms with exponential dependence in $e^{k_3(z-R)}$ 
and $e^{k_4(z-R)}$ associated with the very rapidly decaying and 
oscillating Ekman solutions. 

We now compare the relative amplitudes of all of 
these terms as a function of the parameters of the system. The amplitude 
of the rapidly decaying component of the solution arising from 
azimuthal forcing only is 
\begin{equation}
d_{\rm E}^2  k_1^2 \frac{U_0}{k_2} \left| \frac{k_3k_4}{k_3-k_4} \right| 
\simeq  \sqrt{\frac{8 E_\nu}{Pr Bu} } \frac{D}{R} U_0
\end{equation}
using the values of $k_i$ derived in equations (\ref{eq:kis}) and 
(\ref{eq:k2}), and where
\begin{equation}
E_\nu = \frac{\nu}{R^2 \Omega } = \frac{2 d_{\rm E}^2}{R^2}
\end{equation}
is the Ekman number. Similarly, the slowly decaying component 
of the solution arising from azimuthal forcing only has an amplitude
\begin{equation}
d_{\rm E}^2 k_1^2 U_0 \simeq 2 E_\nu U_0
\end{equation}
Given that $E_\nu \simeq 10^{-16}$ near the base of the convection zone 
for microscopic values of the viscosity, within an Ekman length of the 
boundary both components of the solution are negligible.

Similarly with the other forcing contributions, we conclude that 
\begin{itemize}
\item latitudinal forcing drives meridional flows with a 
rapidly decaying component which has an amplitude $V_0$ 
(with no dependence on any of the other parameters of the system),  
while the slowly decaying component has an amplitude proportional to 
$E_\nu V_0$. Thus, in agreement with the study of GM04, we find that within 
a few 
Ekman lengths, both are negligible for microscopic values of the viscosity.
\item radial forcing drives meridional flows with a 
rapidly decaying component which has an amplitude proportional to 
$W_0/\sqrt{E_\nu}$, while the slowly decaying component has an 
amplitude proportional to $\sqrt{E_\nu} W_0$. Thus, in this particular 
case we find that beyond a few Ekman 
lengths the slow mode retains a non-negligible amplitude of about $10^{-8}$ 
times the imposed velocity. Indeed, for imposed meter per second flow 
velocities, the flows near the top of the radiative zone could then be of the 
order of $\sqrt{E_\nu} W_0 \simeq 10^{-6}$ centimeters per second, with an 
overall turnover time of the order of $R/\sqrt{E_\nu} W_0 \simeq 1$ Gyr. 
While very slow, this can still 
provide mixing in the tachocline on the stellar evolution and/or 
gravitational settling timescale. 
\end{itemize}

\subsection{Thermal boundary conditions of type 2: no net perturbed heat flux through the boundary}
\label{subsec:case2} 

Dropping the assumption that the convection zone is a perfectly 
conducting fluid, we require instead that  
$ \overline{h} w = c_{\rm p} \kappa \partial \theta / \partial z$. 
which is equivalent to $\overline{T} w = \kappa \partial \theta / \partial z$.

We can rewrite this new thermal boundary condition into 
the linear systems $M A = C$ and $MB = E$ where $M$, $A$, $B$ and $C$ have 
already been defined, and
\begin{equation}
E = \begin{pmatrix} 0 \\ V_0d_{\rm E}^{-2} \\ W_0 d_{\rm E}^{-2} k_1^{-1} \\ 
- W_0 k_2^2 H_{\Theta} / d_{\rm E}^{2} k_1 \end{pmatrix}  \mbox{   ,   } 
\end{equation}
where $H_{\Theta}$ is the potential temperature scaleheight 
\begin{equation}
H_{\Theta} = \frac{\overline{T}}{\beta} = \frac{g}{N^2}  \mbox{   .   } 
\end{equation}
Thus the $\{a_i\}$ coefficients are the same as in the previous section, while
 in the limit where $k_2 \ll k_1 \ll |k_3|,|k_4|$ 
\begin{eqnarray}
b_1 \simeq - \frac{H_{\Theta} k_2^2}{d_{\rm E}^2 k_1^3} W_0 e^{-k_1R}   \mbox{   ,   } 
\nonumber \\
b_3 \simeq \left( - \frac{V_0}{k_3k_4} +  \frac{H_{\Theta} k_2^2}{k_1^3}W_0  
\right) \frac{e^{-k_2R}}{d_{\rm E}^2}   \mbox{   ,   }  \nonumber \\
b_5 \simeq - \left[ V_0 + W_0 \frac{k_4}{k_1} \left(H_{\Theta}k_2 - 1\right) 
\right]  \frac{e^{-k_3R}}{k_3(k_3-k_4)d_{\rm E}^2}  \mbox{   ,   }  \nonumber \\
b_7 \simeq - \left[V_0 + W_0 \frac{k_3}{k_1} \left(H_{\Theta}k_2 - 1\right)  
\right] \frac{ e^{-k_4R}}{ k_4(k_4-k_3) d_{\rm E}^2 } \mbox{   ,   } 
\end{eqnarray}
so that 
\begin{eqnarray}
v(x,y,z) &\simeq&  \left[ - d_{\rm E}^2 k_1^2 U_0 e^{k_2(z-R)} 
- d_{\rm E}^2 k_1^2 \frac{U_0}{k_2} \frac{k_4k_3}{k_3-k_4} e^{k_3(z-R)}  
\right. \nonumber \\ &-& \left. d_{\rm E}^2 k_1^2 \frac{U_0}{k_2} 
\frac{k_3k_4}{k_4-k_3}  e^{k_4(z-R)}  \right] \cos\left( \frac{2 y }{R} \right)
 \nonumber \\ 
&+& \left[ \left( - \frac{V_0 k_1^2}{k_3k_4} 
+ \frac{H_{\Theta} k_2^2}{k_1}W_0  \right)  e^{k_2(z-R)} \right.  \nonumber \\ 
&+& \left( V_0 + W_0\frac{k_4}{k_1} \left(H_{\Theta}k_2 - 1\right) \right)  
\frac{k_3e^{k_3(z-R)}}{k_3-k_4}  \\ &+& \left. 
\left(V_0  + W_0 \frac{k_3}{k_1} \left(H_{\Theta}k_2 - 1\right)  \right)  
\frac{k_4e^{k_4(z-R)}}{k_4-k_3}  \right] \sin\left( \frac{2 y }{R} \right) \mbox{   .   } \nonumber 
\label{eq:vcase2}
\end{eqnarray}

For this new set of boundary conditions, azimuthal forcing and latitudinal 
forcing yield the same solutions as in the previous section, 
while radial forcing drives meridional flows with a rapidly 
decaying component which has an amplitude proportional to 
$W_0 (H_{\Theta}k_2 - 1)/\sqrt{E_\nu}$, while the slowly decaying component 
has an amplitude proportional to $(H_{\Theta}/R) (R/D)^2 Pr Bu W_0$.
Note how in this case the slowly decaying component of the flow has 
an amplitude which is independent of the background viscosity $\nu$.

\subsection{Discussion of the solutions}
\label{subsec:disc}

Using this Cartesian geometry model, we have shown that there exist solutions
for meridional flows in the radiative interior with significant amplitude 
throughout. These flows can only be present if driven by direct pumping into 
and out of the radiative zone, that is, when $w_{\rm cz}(y)$ has a 
significant amplitude on the boundary. Our study also shows that the actual 
amplitude of the flows within the radiative zone depends sensitively on the 
thermal boundary conditions used. While it is not clear which set of boundary 
conditions actually accurately represents the true convective-radiative
 interface, it is not implausible that they may lie somewhere in between 
the two extreme cases studied. Thus it is also not implausible that there
may be significant penetration of the convective zone flows into the 
radiative zone, with typical turnover times shorter than $R/\sqrt{E_\nu} W_0$. 

Our results extend the work of GM04, 
and naturally recover their solutions in the limit where only latitudinal 
forcing is taken into account. Note that GM04 do not specifically require that
$w$ be null on the convective-radiative interface, and argue that any
spatially varying latitudinal flow drives radial flows into and out of the 
boundary to guarantee mass conservation. This is indeed correct, but one must 
keep in mind that $\partial v/\partial y = - \partial w /\partial z$ only
relates latitudinal variations in $v$ to the radial derivative of $w$. Since
the derivative is dominated at the base of the convection zone by rapid 
variation on the Ekman scale, the mass conservation equation in fact requires 
that the radial flows in the GM04 solution have an amplitude of the order of 
$W_0 = V_0 d_{\rm E}/R$. It is therefore easy to see why their solution 
yields negligible velocities for the slowly varying component of the flow
penetrating into the radiative zone.

Finally, it is crucial to note that since the momentum and energy equations 
in this study (and the preceding ones by GM04 and McIntyre, 2007) 
have been linearized, the predicted amplitudes of the flows are 
linearly dependent on the imposed flow velocities. In reality, this 
will only be true in the limit where the typical amplitudes of the nonlinear
advection terms ($\bu \cdot \nabla \bu$ for the momentum equation, 
and $\bu \cdot \nabla \theta$ for the energy equation) are indeed much 
smaller than the corresponding linear terms 
($2\Omega \times \bu$ for the momentum equation and $\beta w$ for the 
energy equation). The nonlinear terms in the momentum equation for 
example guarantee that no unphysical counter-rotation is allowed. The 
linearized equations on the other hand scale arbitrarily with the imposed
boundary conditions and do allow counter-rotation for certain input 
parameters, e.g. if the 
imposed differential rotation profile $u_{\rm cz}(y)$ is large enough or 
if the meridional flows are too rapid (see Figure \ref{fig:nm8} for instance).
Similarly, the nonlinear terms in the 
energy equation guarantee that the actual turnover time of the meridional flows
cannot be lower than the local thermal diffusion time (over the depth 
considered), while the linearized equations allow for any values of the 
turnover time provided the imposed boundary flow velocities are high enough.

From this linearized model, we can therefore say with reasonable confidence 
that meridional flows
can indeed penetrate into the radiative zone, provided their turnover time 
(typically estimated as $v(d)/d$ where $d = R-z$ is the depth considered) 
is longer than the local thermal diffusion time (estimated as $d^2/\kappa$).

\section{A reduced Cartesian model}
\label{sec:toy}

One of the remaining issues that needs to be addressed is that of lower 
boundary. In the previous section, the high-dimensionality of the solution 
space made it difficult to consider the more realistic situation of a 
bottom boundary located at $z=0$. While the Ekman solutions (associated 
with $k_3$ and $k_4$) decay inward/downward so quickly that the presence 
or absence of a lower boundary cannot affect them, the slowly varying 
solutions do span the entire radiative region from $z=R$ down to $z=0$;
the location of the lower boundary is expected to have some influence
on their amplitude.

Here, we therefore focus on studying the slowly varying solutions only 
by considering a system in which the Ekman flows are filtered out. This
is another way of reducing the dimensionality of the solution space, and 
enables us to compare the solutions in a semi-infinite domain to the solutions 
in a finite domain.

\subsection{Reduced model equations}

We consider the system 
\begin{eqnarray}
\frac{\ptl v}{\ptl y} + \frac{\ptl w}{\ptl z} = 0  \mbox{   ,   } \nonumber \\
\frac{\ptl p}{\ptl z} = \alpha g \theta \mbox{   ,   }  \nonumber \\
2 \Omega u = -\frac{\ptl p}{\ptl y}  \mbox{   ,   } \nonumber \\
-2 \Omega v = \nu \left( \frac{\ptl^2 u}{\ptl y^2} + \frac{\ptl^2 u}{\ptl z^2} 
\right)  \mbox{   ,   } \nonumber \\
\beta w = \kappa \left( \frac{\ptl^2 \theta}{\ptl y^2} 
+  \frac{\ptl^2 \theta}{\ptl z^2} \right) \mbox{   ,   } 
\label{eq:maineqs2}
\end{eqnarray}
in which the viscous diffusion terms within the latitudinal component of
 the momentum equation has now been removed. This simplification is 
consistent with the assumptions of geostrophic equilibrium and, as we now 
prove, effectively filters out the Ekman flows. The system can be reduced 
to a single partial differential equation for $u$, for instance, as
\begin{equation}
\frac{\partial^4 u}{\ptl z^4} + \left( 1 + \frac{d_{\rm E}^4}{d_{\rm BD}^4} 
\right) \frac{\partial^2 u}{\partial z^2 \partial y^2} 
+ \frac{d_{\rm E}^4}{d_{\rm BD}^4} \frac{\ptl^4 u}{\ptl y^4}= 0 \mbox{   .   } 
\end{equation}

Seeking exponential solutions in $z$ with periodic behavior in $y$ as before 
yields the characteristic equation:
\begin{equation}
\left(k^2 - \frac{4n^2}{R^2}  \right) \left( k^2 - \frac{4 n^2}{R^2} 
\frac{d_{\rm E}^4}{d_{\rm BD}^4} \right) = 0 \mbox{   ,   } 
\end{equation}
with solutions $\pm K_1$ and $\pm K_2$ with 
\begin{eqnarray}
K_1 = \frac{2n}{R} = k_1  \mbox{   and   }  \nonumber \\
K_2 = \frac{2n}{R} \left( \frac{d_{\rm E}^2}{d_{\rm BD}^2} \right) = k_2  \mbox{   .   } 
\end{eqnarray}
Thus we recover only the slowly varying solutions of the previous section.

\subsection{General solutions}
 
The flow solution to the above system for fixed latitudinal wavenumber $n$ is 
\begin{eqnarray}
u(x,y,z) &=& \left(A_1 e^{K_1z} + A_2 e^{-K_1z} + A_3 e^{K_2z} + A_4 e^{-K_2z} 
\right) \cos\left(\frac{2ny}{R} \right) \nonumber \\ 
&+& \left(B_1 e^{K_1z} + B_2 e^{-K_1z} + B_3 e^{K_2z} + B_4 e^{-K_2z} \right) 
\sin\left(\frac{2ny}{R} \right)
\end{eqnarray}
which also yields
\begin{eqnarray}
v(x,y,z) &=&
d_{\rm E}^2 \left( K_1^2 - K_2^2 \right) \left( A_3 e^{K_2z} + A_4 e^{-K_2z}  
\right) \cos\left(\frac{2ny}{R} \right) \nonumber \\ 
&+&  d_{\rm E}^2 \left( K_1^2 - K_2^2 \right) 
\left( B_3 e^{K_2z} + B_4 e^{-K_2z}  \right) \sin\left(\frac{2ny}{R} \right)
\end{eqnarray}
and 
\begin{eqnarray}
w(x,y,z) &=&  d_{\rm E}^2 \frac{K_1}{K_2}\left(K_1^2 - K_2^2 \right) 
\left( A_3 e^{K_2z} - A_4 e^{-K_2z}  \right) \sin\left(\frac{2ny}{R} \right) 
\nonumber \\ 
&-&  d_{\rm E}^2 \frac{K_1}{K_2}\left(K_1^2 - K_2^2 \right)  
\left( B_3 e^{K_2z} - B_4 e^{-K_2z}  \right) \cos\left(\frac{2ny}{R} \right)
\end{eqnarray}

\subsection{Boundary conditions}

The governing equations considered form what is apparently a 6$^{\rm
th}$ order system in the $z$ direction, and require, in principle, 6
boundary conditions split between the top and bottom
boundaries. Naturally, one would like to impose boundary conditions on
$u$, $w$ and $\theta$ (one at the top, one at the bottom for each
variable). Note that since the viscous terms have been
neglected in the equations for hydrostatic and geostrophic equilibrium
respectively, one cannot impose a boundary condition on $v$: the
equations contain no stresses that could transfer the boundary
information to the rest of the system. However, we see that 
combining these equations only yields a 4$^{\rm th}$ order partial 
differential equation for $u$. This implies that
the 6 selected boundary conditions must somehow be redundant,
otherwise there will be no solution to the system. In what follows, we
therefore only select boundary conditions for $u$ and $w$ at the top
and bottom boundaries. Near the top boundary, we consider as before
\begin{eqnarray}
u(x,y,R) = u_{\rm cz}(y) \mbox{   ,   } \nonumber \\
w(x,y,R) = w_{\rm cz}(y) \mbox{   .   } 
\end{eqnarray}
When considering a bottom boundary at $z=0$, we choose impermeable boundary conditions for $w$, and stress-free boundary conditions for $u$ so that
\begin{eqnarray}
\left. \frac{\partial u}{\partial z} \right|_{(x,y,0)} = 0\mbox{   ,   }  \nonumber \\
w(x,y,0) = 0\mbox{   .   } 
\end{eqnarray}

\subsection{Solutions for a semi-infinite domain}

When the bottom boundary is near $-\infty$, we find that 
\begin{eqnarray}
A_1 = -U_0 e^{-K_1 R} \mbox{   ,   } \nonumber \\
A_2 = A_3 = A_4 = 0 \mbox{   ,   } \nonumber \\
B_1 = - \frac{K_2}{K_1(K_1^2-K_2^2)} \frac{W_0}{d_{\rm E}^2} e^{-K_1R} 
\mbox{   ,   } \nonumber \\
B_2 = B_4 = 0 \mbox{   ,   } \nonumber \\
B_3 = \frac{K_2}{K_1(K_1^2-K_2^2)} \frac{W_0}{d_{\rm E}^2} e^{-K_2R}\mbox{   ,   } 
\end{eqnarray}
so that 
\begin{equation}
v(x,y,z) = \frac{K_2}{K_1} W_0 e^{K_2(z-R)} \sin(K_1 y)\mbox{   ,   } 
\label{eq:semiinf}
\end{equation}
which illustrates again how radial forcing can yield non-zero flow 
amplitudes penetrating deeply into the radiative zone. It is important 
to note, however, that the predicted flow amplitude is different from that 
found in \S\ref{sec:cart}. This might be attributed to the fact that when 
viscous effects are taken into account, a fraction of the flow penetrating 
into the radiative zone is deflected into the very shallow Ekman layers. 

\subsection{Solutions for a finite domain}

When the bottom boundary is located at $z=0$, we find that 
\begin{eqnarray}
A_1 &=& A_2 = - \frac{U_0}{2 \cosh(K_1 R)} \mbox{   ,   } \nonumber \\
A_3 &=& A_4 = 0 \mbox{   ,   } \nonumber \\
B_1 &=& B_2 = - \frac{K_2}{K_1(K_1^2-K_2^2)} \frac{W_0}{d_{\rm E}^2} 
\frac{\cosh(K_2R)}{2 \sinh(K_2R) \cosh(K_1R)} \nonumber \\ 
&\simeq& - \frac{W_0}{2d_{\rm E}^2 K_1^2} \frac{1}{K_1R \cosh(K_1R)} \mbox{   ,   } 
\nonumber \\
B_3 &=& B_4 = \frac{K_2}{K_1(K_1^2-K_2^2)} \frac{W_0}{d_{\rm E}^2} 
\frac{1}{2 \sinh(K_2R)} \nonumber \\ 
&\simeq& \frac{W_0}{2d_{\rm E}^2 K_1^2} \frac{1}{K_1R}\mbox{   ,   } 
\end{eqnarray}
implying
\begin{eqnarray}
v(x,y,z) &=& \frac{W_0}{\sinh(K_2R)} \frac{K_2}{K_1} \cosh(K_2z) \sin(K_1 y) \nonumber \\
 &\simeq& \frac{W_0}{K_1R} \cosh(K_2z) \sin(K_1 y) \mbox{   .   } 
\label{eq:finite}
\end{eqnarray}

\subsection{Consequences}
\label{subsec:cons}

Comparing the expressions in equations (\ref{eq:semiinf}) and 
(\ref{eq:finite}), we find that the slowly varying component of the  
meridional flow in the case of a finite domain has an amplitude that is 
$1/K_2 R$ times that of the semi-infinite domain case.
The difference between the predicted amplitudes for the two geometrical 
systems (finite and semi-infinite domains) can easily be understood in the 
light of the fact that the exponential solutions associated with 
$K_1$ and $K_2$ decay on much longer lengthscales than $R$. 

Extrapolating this result to the full Cartesian problem studied in 
\S\ref{sec:cart}, 
we therefore predict that the slowly varying components of the meridional 
flows (associated with the $k_1$ and $k_2$ wavenumbers) should in fact have 
an amplitude that is $1/ k_2 R $ times that given in 
(\ref{eq:vcase1}) and (\ref{eq:vcase2}); 
the Ekman components on the other hand decay so rapidly that their amplitude
should not be influenced by the presence of a lower boundary. We now 
revise our estimates of \S\ref{subsec:case1} and \S\ref{subsec:case2}
to predict that in the case of type 1 boundary conditions 
($\partial \theta /\partial z = 0$ at the upper boundary) then
\begin{itemize}
\item azimuthal forcing leads to meridional flows with a rapidly decaying 
component with an amplitude proportional to $\sqrt{E_\nu/Pr Bu} U_0 $, 
while the slowly decaying component has an amplitude proportional to 
$E_\nu U_0/\sqrt{Pr Bu}$. 
\item latitudinal forcing leads to meridional flows with a rapidly decaying 
component with an amplitude proportional to $V_0$, while the slowly decaying 
component has an amplitude proportional to $E_\nu V_0/\sqrt{Pr Bu}$. 
\item radial forcing leads to meridional flows  with a rapidly decaying 
component with an amplitude proportional to $W_0/\sqrt{E_\nu}$, while the 
slowly decaying component has an amplitude proportional to 
$\sqrt{E_\nu/PrBu} W_0$. 
\end{itemize}
In the case of type 2 boundary conditions (where $\kappa \partial \theta /\partial z = \overline{T} w $ at the upper boundary) 
the flow velocities predicted in the case of azimuthal and 
latitudinal forcing are the same, while radial forcing drives flows 
with a rapidly decaying component with an amplitude proportional to 
$W_0 (H_{\Theta}k_2 - 1)/\sqrt{E_\nu}$ and a slowly decaying 
component which has an amplitude proportional 
to $(H_{\Theta}/R) \sqrt{Pr Bu} W_0$. 

\section{Numerical solutions in a spherical shell}
\label{sec:num}

\subsection{Model setup}

To complete this axisymmetric study and illustrate our analytical 
results numerically, we perform 
simulations of flows in the solar radiative zone subject to various boundary 
conditions. The governing equations for the numerical model are derived by 
perturbing the spherically symmetric equations of stellar structure, moving 
to a rotating frame of reference and assuming that the velocity perturbations 
and thermodynamical perturbations to the background spherically symmetric 
state are small enough for linearization to be appropriate. Then
\begin{eqnarray}
&& 2\overline{\rho} \overline{\bO} \times \bu   =
- \grad \tilde{p} - \tilde{\rho} \grad \overline{\Phi}  
+  f \div \Pi \mbox{   ,   } \nonumber  \\
&&  \div(\overline{\rho} \bu) = 0 \mbox{   ,   }\nonumber  \\
&& \overline{\rho} \overline{T} \bu\cdot \grad \overline{s}  
= f \div( \overline{k} \grad \tilde{T}) \mbox{   ,   } \nonumber  \\
&& \frac{\tilde{p}}{\overline{p}} = \frac{\tilde{\rho}}{\overline{\rho}} +  \frac{\tilde{T}}{\overline{T}} \mbox{   ,   }
\label{eq:global}
\end{eqnarray}
where $\bu = (u_r, u_\theta, u_\phi)$ is the velocity field in a 
frame rotating with angular velocity 
$\overline{\bO}$, $\rho$, $T$, $s$ and $p$ are the standard
thermodynamical variables, $k = \overline{\rho} c_{\rm p} \kappa $ 
is the thermal conductivity, $\Pi$ 
is the viscous stress tensor (which depends on the viscosity $\nu$) and 
$\Phi$ is the gravitational potential. Quantities denoted with bars are 
background quantities, taken from the standard solar model of 
Christensen-Dalsgaard {\it et al.} (1991), while the temperature, pressure 
and density perturbations are denotes with tildes. 

This system is fully consistent with the Cartesian
model equations presented in \S\ref{sec:cart}, with the added 
sophistication of the perfect gas equation of state and the anelastic 
approximation instead of the simpler 
Boussinesq approximation. This modification is added so that the model 
equations are consistent with the level of approximation used, but does 
not affect the nature of the solutions. The centrifugal force associated
with the rotation of the background state 
$\overline{\bO} \times \overline{\bO} \times \br$ 
has been removed to suppress global Eddington-Sweet circulations, which are 
known to be of very small amplitude, but would otherwise play an 
important role in this steady-state calculation (see the work of Garaud, 
2002, for comparison).  Perturbations in the gravitational potential
are neglected in accordance with Cowling's approximation.

Note how
both diffusion terms have been multiplied by the same factor $f$. Since the 
typical Ekman number in the Sun (just below the base of the convection zone) is
of the order of $10^{-16}$, a unit value of $f$ would lead to Ekman layers 
about $10^{-8}$ times size of the domain; the 
numerical method used is unable to resolve them. Using values of 
$f$ of the order of $10^7-10^9$ instead inflates the Ekman layers 
artificially to $10^{-4}$ times the size of the domain or 
larger, which can then be fully resolved. As an added bonus, varying $f$ 
provides an easy way of varying the effective viscosity and thermal 
conductivity without changing the Prandtl number. As a result, the estimated
values of $k_1$ and $k_2$ are unchanged from the solar value
(since $k_2$ only depends on the Prandtl number), while $k_3$ and $k_4$ are a
factor of $f^{1/2}$ smaller. As long as $f \ll 10^{15}$, the hierarchy  
$k_2 \ll k_1 \ll |k_3|, |k_4|$ is respected.

The numerical method of solution is based on the expansion of the governing
equations onto a spherical coordinate system $(r,\theta,\phi)$, followed
by their projection onto Chebishev polynomials
$T_n(\cos\theta)$, and finally, solution of the resulting ODE system in $r$ 
using a Newton-Raphson-Kantorovich algorithm. For more detail, see 
Garaud (2001). The computational domain is limited 
to the region of the radiative zone within $r \in [0.02,0.7]r_\odot$. The
solution is found to be reasonably 
insensitive to the position of the lower boundary for this range of 
parameters.

In order to study the dependence of the amplitude and depth of penetration 
of the flows on the Ekman number (i.e. on $k_3$ and $k_4$), 
we perform a series of simulations with $f$ 
ranging from $10^7$ to $10^{9}$. In order to verify the dependence of the
amplitude of the solutions on $k_2$, we perform similar calculations with 
a hypothetical Sun where the thermal conductivity in the system of equations 
(\ref{eq:global}) is uniformly multiplied 
by a factor of four throughout the interior, resulting in a Prandtl number 
artificially decreased by a factor of four compared with its solar value. 
In that case, $k_2$ is two times smaller than in the case of the solar
value of the Prandtl number.

In order to study in detail the effects of the boundary conditions on the 
solutions, we study separately 
three cases where forcing is respectively in the azimuthal, latitudinal 
and radial directions only. In the case of radial forcing, 
we consider the two boundary conditions studied in \S\ref{sec:cart}, namely 
$\partial \tilde{T}/\partial r$ is null on the boundary, 
and $\kappa \partial \tilde{T}/\partial r = \overline{T} u_r$. 

The details of the expressions for the boundary conditions in each case 
are given below.

\subsection{Boundary conditions.}
\label{subsec:numbc}

In all of the simulations performed, the lower boundary conditions at 
$r = 0.02r_\sun$ are the following: 
\begin{itemize}
\item impermeable condition on the radial velocity,
\item stress-free conditions on the tangential velocities $u_\theta$ and 
$u_\phi$,
\item conducting condition on the temperature: we assume that the domain 
within 0.02$r_\sun$ is a conducting solid sphere, so that the temperature 
perturbations satisfy $\nabla^2 \tilde{T} = 0$ within, and are regular at 
$r=0$. We solve this equation with the requirement that 
$\tilde{T} \rightarrow 0$ as $r \rightarrow 0$, and derive a matching
 condition with the temperature fluctuations at the interface.
\end{itemize}

Near the upper boundary (at $r = 0.7r_\sun$), we consider the following cases:
\begin{itemize}
\item azimuthal forcing only: in this case, we set $u_r = u_\theta = 0$ at 
the boundary (assuming no-slip conditions). Following the results 
from helioseismology, we set $u_\phi = 0.7r_\sun \sin\theta \tilde{\Omega}$ 
with $\tilde{\Omega} = \Omega_{\rm eq} (1-a\cos^2\theta- b\cos^4\theta) 
- \overline{\Omega}$, where $\overline{\Omega} = \Omega_{\rm eq} 
( 1 - a/5 - 3b/35)$ (Gilman, Morrow \& DeLuca, 1989), $a = b = 0.15$ and 
$\Omega_{\rm eq} = 2.9 \times 10^{-6}$s$^{-1}$. The temperature boundary 
condition is $\partial \tilde{T} /\partial r$ = 0.
\item latitudinal forcing only: in this case we set $u_r = u_\phi = 0$ at the 
boundary, and $u_\theta = V_0 \sin\theta \cos\theta$ with $V_0 = 1$ m/s. 
Note that $u_\phi = 0$ is guaranteed by setting $a=b=0$, in which case the 
background angular velocity is $\overline{\Omega} = \Omega_{\rm eq} 
= 2.9 \times 10^{-6}$s$^{-1}$. The temperature boundary condition is 
$\partial \tilde{T} /\partial r = 0$.
\item radial forcing only: in this case we set $u_\theta = u_\phi = 0$ at 
the boundary and $u_r = U_0 (1-3\cos^2\theta)$ with $U_0 = 1$cm/s. 
This expression satisfies the global conservation of mass. In this last 
case, two temperature boundary conditions are explored as in \S\ref{sec:cart}, 
either $\partial \tilde{T}/\partial r = 0$ (type 1) or 
$\kappa \partial \tilde{T}/\partial r  =\overline{T} u_r$ (type 2). 
\end{itemize}
Note that since the governing equations are linear, the amplitudes of 
each component of the flow defined by $(U_0,V_0,W_0)$ at the boundary can 
be chosen arbitrarily. Here, they were selected as what may be plausible flow 
velocities in the lower regions of the convection zone.

\subsection{Results}

This section summarizes the numerical results for the various simulations 
performed. 

In order to compare quantitatively the numerical solutions with the 
Cartesian model predictions we study the variation of the latitudinal 
component of the meridional flow $u_\theta$, both near and far from the 
boundary, at a fixed latitude fairly close to the polar regions. We 
select a high latitude of about $80^\circ$ since the Cartesian model 
applies best to systems in which the rotation axis and gravity are nearly 
aligned. 

\begin{figure}
\epsscale{2}
\plotone{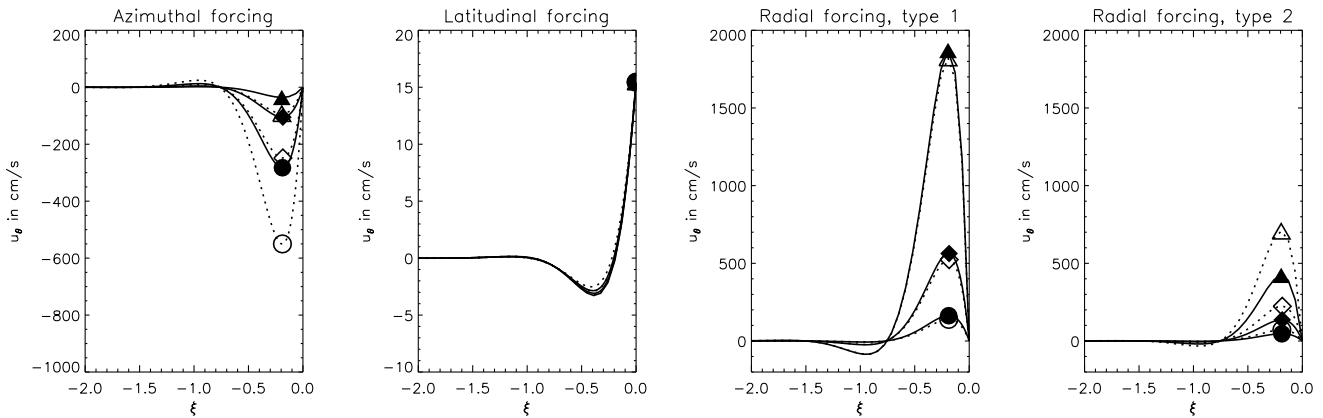}
\caption{Latitudinal velocity at a latitude of 80$^{\circ}$ 
as a function of re-scaled 
depth below the convection zone $\xi = (r - 0.7 r_\sun)/d_{\rm E}$ for 
the four types of boundary conditions described in \S\ref{subsec:numbc}. 
In each of the four plots, the solid lines (and solid symbols) correspond 
to simulations with a solar value of the Prandtl number, while the dotted 
lines (and open symbols) correspond to simulations with a Prandtl number 
artificially reduced by a factor of four. The symbols identify the value 
of $f$ used in the simulations: a circle corresponds to $f=10^9$, a diamond 
to $f= 10^8$ and a triangle to $f = 10^7$. For comparison, the forcing 
velocities at the latitude considered are about $u_{\rm cz}(80^{\circ}) 
= 5200$cm/s, $v_{\rm cz}(80^{\circ}) = 17$cm/s and $w_{\rm cz}(80^{\circ})
= 0.9$ cm/s.}
\label{fig:rapidsol}
\end{figure}

In Figure \ref{fig:rapidsol} we first focus on the region 
close to the upper boundary in order to single out the Ekman solution. 
Each plot corresponds to one of the four types of boundary conditions studied, 
and shows $u_\theta(r)$ at a latitude of about $80^\circ$. For clarity, 
the depth below the convection zone is rescaled with the Ekman depth 
$d_{\rm E}$:  thus, in each 
of the plots $\xi = (r - 0.7 r_\sun)/d_{\rm E}$. The 
results for the solar value of the Prandtl number are shown in solid lines, 
and those corresponding to the lower value of the Prandtl number (equivalently,
a value of $\kappa$ that is four times solar) are shown 
in the dotted lines. In each case, three runs are presented with 
$f = 10^7$, $10^8$, and $10^9$ respectively and can be identified with the 
symbols. It is immediately obvious from observing all of these plots that 
there is indeed a component of the solution which decays 
and oscillates rapidly with depth below the convection zone on an Ekman 
lengthscale.
From the conclusions of \S\ref{subsec:cons}, we expect the amplitude of this 
rapidly 
decaying solution to scale as $\sqrt{f/Pr}$ in the case of azimuthal forcing, 
to be of order of the forcing velocity for any value of $f$ or $Pr$ in the 
case of the latitudinal forcing, to scale as $1/\sqrt{f}$ for radial forcing 
with type 1 boundary conditions and finally, to scale as
$(H_\Theta k_2 -1)/\sqrt{f}$ for radial forcing with type 2 boundary 
conditions. Note that for solar values of the background state and of 
the Prandtl number, 
using the standard model of Christensen-Dalsgaard {\it et al.} (1991), 
$H_{\Theta}k_2 \simeq 0.8$; for a Prandtl number that is reduced by a factor 
of four, then $H_{\Theta}k_2 \simeq 0.4$. 

\begin{figure}
\epsscale{2}
\plotone{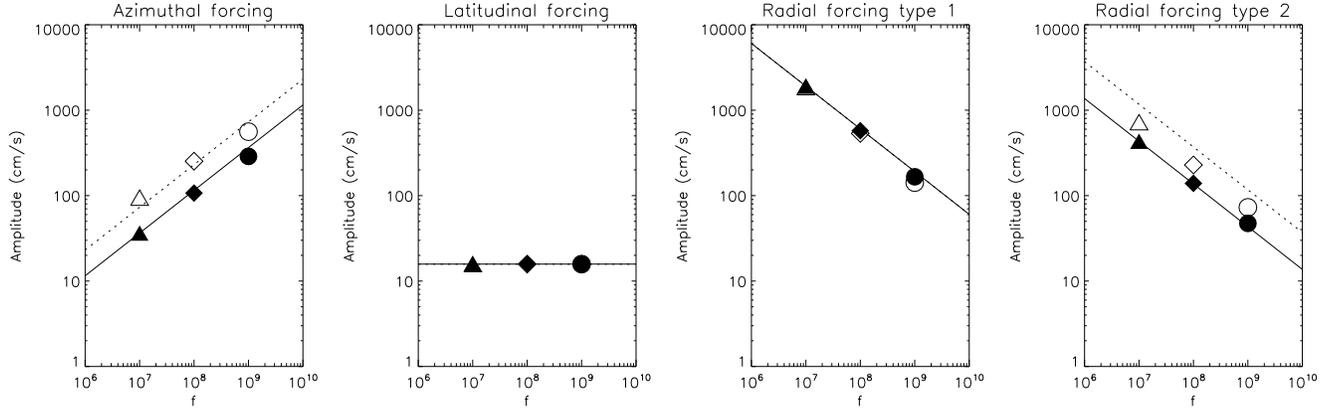}
\caption{Comparison of the scaling of the amplitude
of the rapidly decaying solution with the model predictions for the same
four types of boundary conditions. For each simulation performed (i.e. for 
each type of boundary condition, for each value of $f$ and of the Prandtl 
number considered), the symbols show the maximum value of $|u_\theta(r)|$ 
found numerically at the latitude considered; note that the 
same coding is used as in Figure \ref{fig:rapidsol}. 
The solid and dotted lines shows the predicted scaling 
of this amplitude as a function of $f$ for solar values of the Prandtl number
and a quarter-times solar value of the Prandtl number respectively. The 
amplitude of the solution for solar values of the Prandtl number 
with $f = 10^7$ is used as the reference amplitude.}
\label{fig:rapidscale}
\end{figure}

The comparison between
these predicted scalings and the outcome of the numerical simulations is shown 
in Figure \ref{fig:rapidscale}. Here, the symbols
represent the amplitude of the solution defined as the maximum value achieved 
by $|u_\theta(r)|$ at the selected latitude, for each of the simulations 
performed. The lines show the predicted scalings, using as a reference the 
amplitude of the solution for solar values of the Prandtl number and
with $f = 10^7$. Thus, for example, the solid line in the case of azimuthal 
forcing only is generated by the equation 
\begin{equation}
\hat{A}(f,Pr_\odot) = \hat{A}(10^7,Pr_\odot) \left(\frac{f}{10^7}\right)^{1/2}
\end{equation}
 where $\hat{A}(f,Pr_\odot)$ is the predicted amplitude of the other 
simulations with 
solar values of the Prandtl number, and $\hat{A}(10^7,Pr_\odot)$ 
is the numerically calculated reference amplitude of the solution for 
$f = 10^7$ and 
solar values of the Prandtl number. The dotted line in the same panel is
easily calculated as 
\begin{equation}
\hat{A}(f,0.25 Pr_\odot) = 2 \hat{A}(10^7,Pr_\odot) \left(\frac{f}{10^7}\right)^{1/2}
\end{equation}
The solid and dotted lines in the three other panels are constructed in 
a similar fashion. It is quite clear that the predicted scalings fit the 
numerical solutions very
well, except perhaps in the case of radial forcing with type 2 boundary 
conditions where the fit is only good to within a factor of order unity.
The origin of this discrepancy is not entirely clear, but can be partly 
traced back to non-ideal effects in the equation of state (which affects
the determination of $H_\Theta$) and to geometrical effects 
(which influence the value of $k_2$). 

\begin{figure}
\epsscale{2}
\plotone{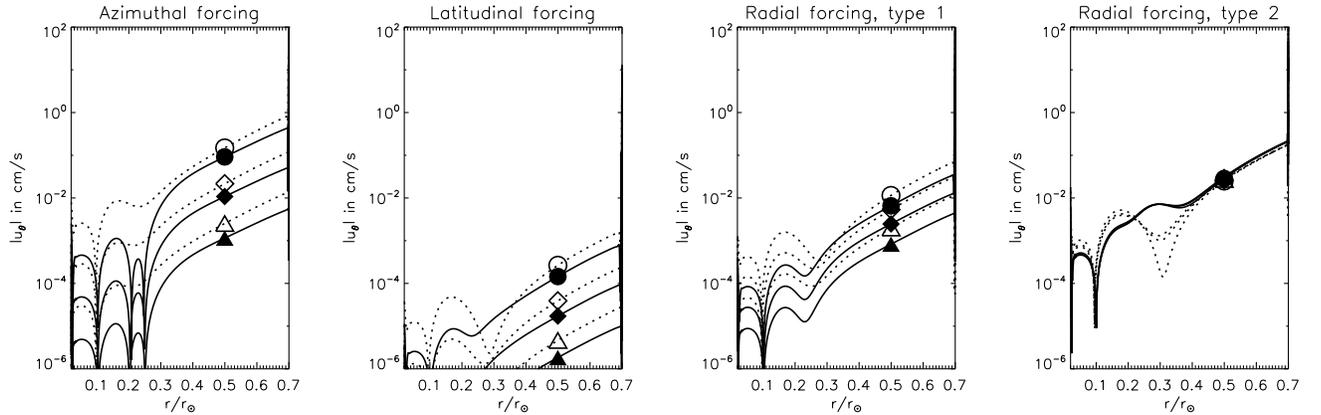}
\caption{Latitudinal velocity of flows below 
the convection zone as a 
function of radius for the four types of boundary conditions described 
in \S\ref{subsec:numbc}. The line and symbol coding are the same as in 
Figure \ref{fig:rapidsol}.}
\label{fig:slowsol}
\end{figure}

\begin{figure}
\epsscale{2}
\plotone{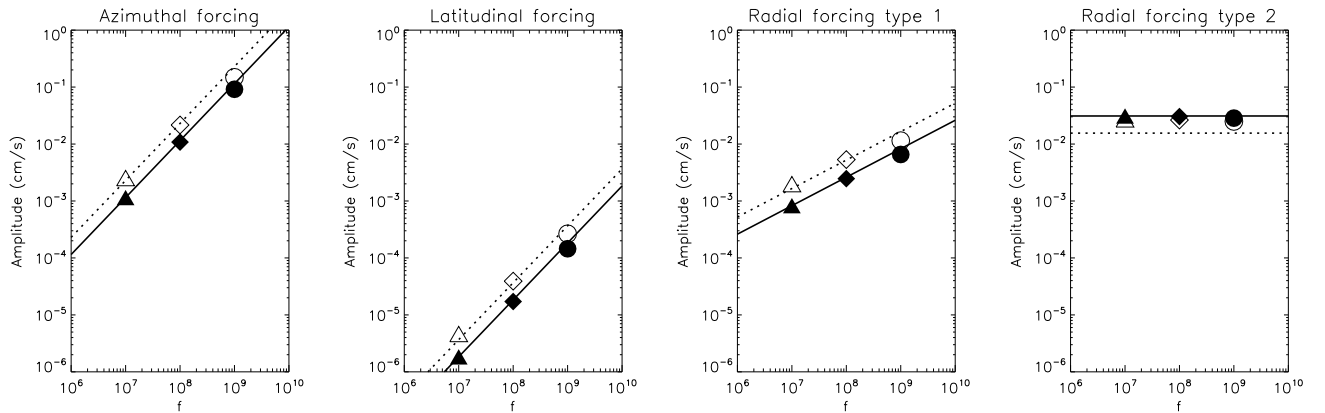}
\caption{Comparison of the scaling of the amplitude
of the slowly decaying solution with the model predictions for the same
four types of boundary conditions. For each simulation performed (i.e. for 
each type of boundary condition, for each value of $f$ and of the Prandtl 
number considered), the symbols show the value of $|u_\theta(0.5r_\odot)|$ 
found numerically at the latitude considered; again, the same coding is used.
The solid and dotted lines shows the predicted scaling 
of this amplitude as a function of $f$ for solar values of the Prandtl number
and a quarter-times solar value of the Prandtl number respectively. The 
amplitude of the solution for solar values of the Prandtl number 
with $f = 10^7$ is used as the reference amplitude. }
\label{fig:slowscale}
\end{figure}

In order to compare the scalings of the slowly decaying component of the
flow with the numerical results, we now move to Figure \ref{fig:slowsol} 
and Figure \ref{fig:slowscale} which focus 
on the behavior of the solution far from the boundary. 
Figure \ref{fig:slowsol} shows $|u_\theta(r)|$ 
at a latitude of about $80^{\circ}$ throughout the interior on a 
log-linear scale. The information about the direction of the 
flow (poleward or equatorward) is lost in this plot, with 
sign reversals appearing as cusps in the curves pointing towards $-\infty$.
Note that it is possible to discern the presence of the Ekman layer 
close to the outer boundary for the larger values of $f$. 

A quick glance at the solutions in the deep radiative interior reveals 
the behavior suggested by the Cartesian solutions: 
different assumptions concerning the boundary conditions yield very different
predicted flow velocities. In particular, 
it can be seen that in the case of radial forcing with type 2 boundary 
conditions (i.e. $\kappa \partial \tilde{T}/\partial r = \overline{T} u_r$)
the flows velocities are more-or-less independent of $f$, or in other
words retain significant amplitudes for any value of the background viscosity.

More quantitatively, from the conclusions of \S\ref{subsec:cons}, 
we expect the amplitude of the 
slowly decaying solution to scale as $f/\sqrt{Pr}$ in the case of azimuthal 
and latitudinal forcing, to scale as $\sqrt{f/Pr}$ for radial forcing 
with $\partial \tilde{T}/\partial r = 0$ and as $(H_\Theta/R)\sqrt{Pr}$ 
in the case of radial forcing with the type 2 
temperature boundary conditions. The comparison between
the predicted scalings and the outcome of the numerical simulations is shown 
in Figure \ref{fig:slowscale}, using the same method
as described earlier in the case of the rapidly decaying component of the 
solution. We can see that, as before, the predicted amplitudes agree very well 
with the numerical solutions, except perhaps in the case of radial forcing with
type 2 boundary conditions where a small discrepancy remains.

We therefore conclude that despite the simplified nature of the analytical
analysis performed in \S\ref{sec:cart} and \S\ref{sec:toy}, the results
obtained robustly predict the scalings of the numerical solutions in full 
spherical geometry.  

\begin{figure}
\epsscale{2}
\plottwo{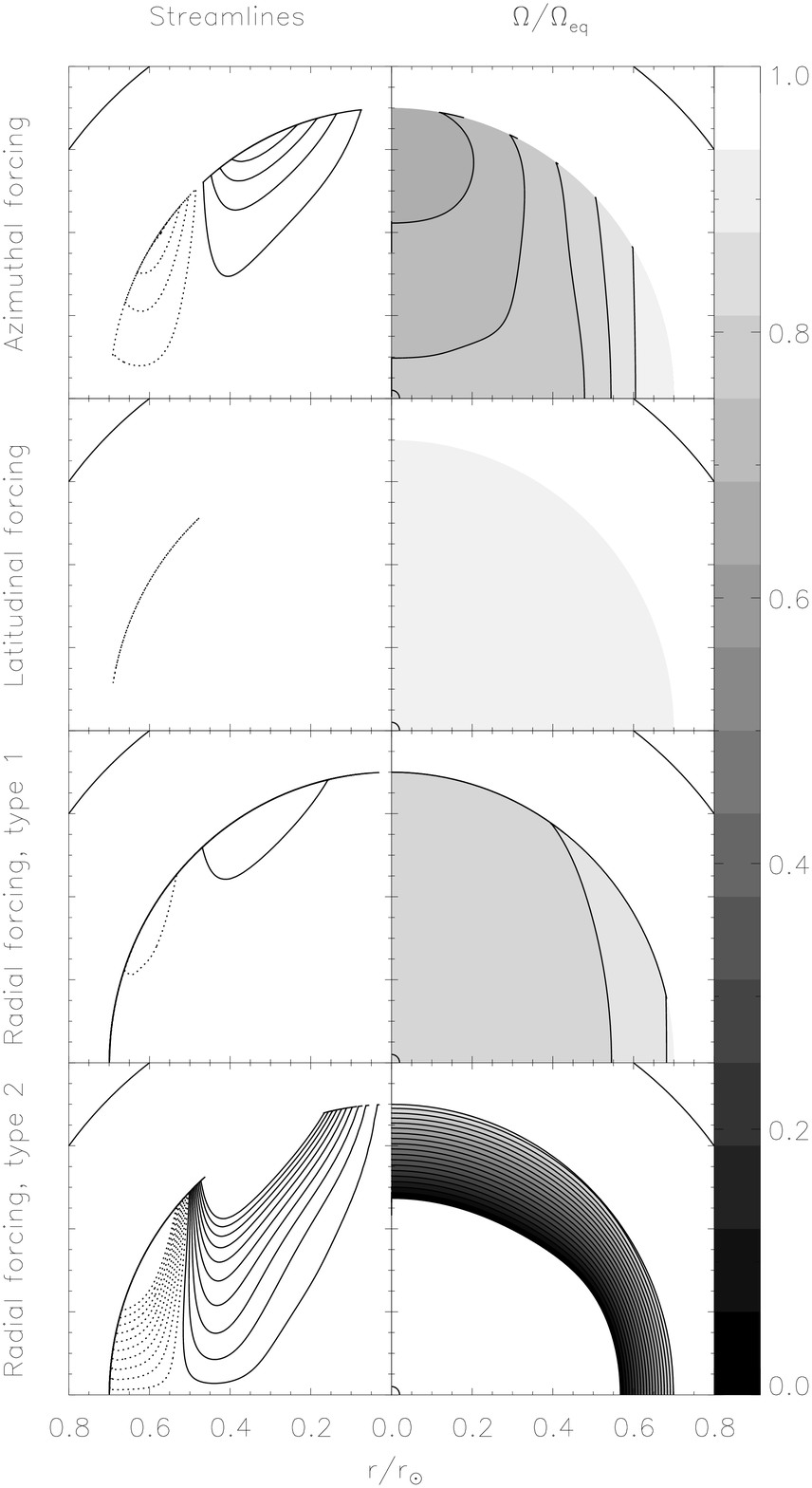}{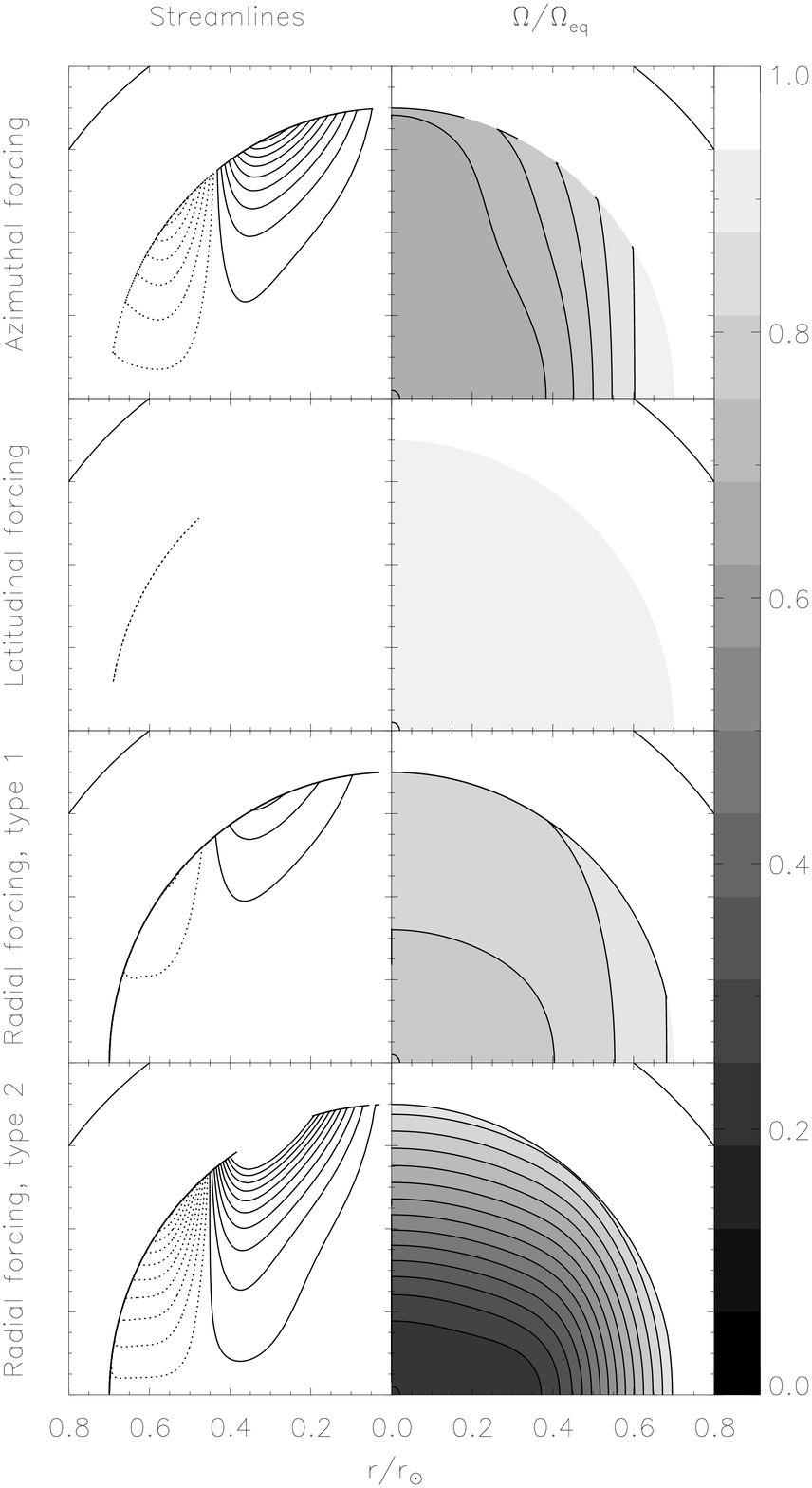}
\caption{Global structure of the flow solutions in the radiative interior 
of the Sun subjected to a series of different boundary conditions at the 
convective-radiative interface (at $r = 0.7 r_\odot$), for fixed value of 
$f = 10^8$. The left-hand-side set of plots shows solutions for a solar 
value of the Prandtl number, 
while the right-hand-side set of plots shows solutions for the case where 
the thermal conductivity is artificially increased by a factor of four. 
In each sets of plots, the streamlines are shown in the left-quadrant; 
the solid lines show clockwise flows, while the 
dotted lines show counter-clockwise flows. The relative angular velocity 
normalized by the equatorial value at the convective-radiative interface 
is shown in the right-quadrant. The ``white'' region in the 
final set of boundary conditions actually corresponds to an unphysical 
counter-rotation. This is an artefact of the linearization of the problem, 
combined with the high flow velocities in the interior, which would not 
occur should the fully nonlinear equations be considered.}
\label{fig:nm8}
\end{figure}

Finally, in order to provide better insight into the actual solutions, we 
show in Figure \ref{fig:nm8} the global 
structure of the numerical results. In the left-hand-side of 
Figure \ref{fig:nm8}, we show results for the solar value of the Prandtl 
number (and therefore solar values of $k_2$ near the convective-radiative 
interface), while in the right-hand-side we show results for a quarter-times
solar value of the Prandtl number. All the plots were generated for 
a fixed value of $f = 10^8$ (and therefore $10^8$-times solar values 
of the viscosity corresponding to 
$10^{-4}$-times solar values of $k_3$ and $k_4$). The left quadrants
show contour-lines of the stream-function, or in other words, streamlines 
of the flow while the right quadrants show the angular velocity profile. 
The four aforementioned 
boundary conditions are explored: azimuthal forcing only, latitudinal 
forcing only, radial forcing only with type 1 boundary conditions 
($\partial \tilde{T}/\partial r = 0$) and radial forcing only with type 2 
boundary conditions 
($\kappa \partial \tilde{T}/\partial r = \overline{T} u_r$).

These numerical results illustrate graphically the behavior predicted by the 
Cartesian model. For example, the streamlines plotted all correspond to 
the same contours of the streamfunction, or in other words 
the same mass flux. It is therefore obvious that the flows driven through 
latitudinal forcing are negligible compared with the three other cases. 
The flows driven by azimuthal forcing appear to be quite strong, 
although this is merely related to the fact that the linear velocity $U_0$ 
corresponding to the imposed (solar) differential rotation is more than 
two orders of magnitude larger than $V_0$ at the same latitudes. The flows
driven by radial forcing are quite strong, in particular given that the 
driving velocity $W_0$ is at least one order of magnitude lower than 
$V_0$ at the same latitude. One can also readily see that the flow 
velocities predicted using type 2 thermal boundary conditions are much 
stronger than in the case of type 1 thermal boundary conditions. 
The right-side quadrants reveal the effect of these meridional flows on 
angular momentum transport in the interior and show
the angular velocity  profile corresponding to the ultimate steady-state 
achieved in the radiative interior should the Sun be left to evolve for many 
flow turnover times under the same applied boundary conditions. Since 
the turnover time can be considerably longer than the age of the Sun 
in most cases, these steady states would never in practise be achieved. 
However, they do reveal two points 
of particular interest. Firstly, they illustrate how sensitive the interior
rotation rate is to the assumed or calculated flow structure, and therefore
also to the assumed convective-radiative interfacial boundary conditions. 
Secondly, they reveal the possibility of unphysical 
counter-rotation permitted by the linearization of the problem 
(see \S\ref{subsec:cons} for a discussion of this effect). This should be taken
as a cautionary warning for any linear study of flows and thermal 
perturbations in the solar radiative interior to always perform a 
self-consistency check of the validity of the linearization.

\section{Discussion and conclusion}
\label{sec:ccl}

We have studied the penetration of global-scale meridional flows
generated in the solar convection zone into the radiative zone below.

We have attacked the problem via a Cartesian model, as others have
done before us, and then
verified our results with more complete numerical modeling in the
correct geometry. We have extended previous work done by GM04 and 
McIntyre (2007) by taking into account the full structure of 
the governing equations and by examining the
effect of boundary conditions on the solutions, or in other words,
allowing a greater range of convective flows to act as sources for the
flows in the radiative interior. 

Within a linear formalism, we confirm that the flow pattern in the 
radiative interior is a linear combination of two types of solutions: 
$(a)$ a solution that decays away from the
convective-radiative interface on a short length-scale related to the
Ekman depth, and $(b)$ a solution that decays away from the interface
on a much longer length-scale associated with the stratification of the 
radiative interior. Our more complete treatment of the problem
provides accurate expressions for the slowly varying solution. 
In addition, we find that the 
{\it  amplitudes} of the rapidly and slowly varying solutions depend 
sensitively on the choice of dynamical and thermal interfacial forcing.

Forcing by azimuthal and
latitudinal shear at the upper boundary lead to both rapidly
varying and slowly varying solutions, but  with amplitudes that are
negligible outside a few Ekman lengths. Forcing by direct radial pumping 
however can generate flows which are significant outside the Ekman 
boundary layer
and indeed are sufficiently slowly-varying that they may maintain
significant flow amplitudes across the whole depth of the radiative
zone. The selected thermal boundary conditions at the convective-radiative 
interface are also found to have significant impact on the interior solution. 


In this linear problem, full solutions
can be built from combinations of the individual solutions obtained for each 
different type of boundary forcing that we studied. Thus in principle
we should be able to determine the complete flow structure within the 
radiative interior. However, the problem lies in the selection of the 
interfacial conditions. We know from helioseismology the amplitude 
of the average azimuthal flows existing near the base of the convection zone, 
but know very little of the nature of the meridional flows or of the thermal 
conditions at that interface. Thus, without a complete model of the whole solar
interior which includes the turbulent solar convection zone we can only 
speculate upon the thermal and dynamical nature of the convective-radiative 
interface. Instead, when studying a broad range of {\it plausible}
boundary conditions we find that significant penetration depths can be 
realized. 

Previous work by GM04 focused on the effects of a latitudinal source 
flow only. Our result is in accordance with theirs for this particular type of 
boundary condition, namely that the flows within the 
radiative interior are limited to an Ekman depth. However, GM04 conclude that 
penetration is {\it always} quenched. Given our analytical and numerical 
results in the case of other applied boundary conditions, we feel that 
this conclusion is too strong. Other plausible models of the interface 
lead to different results concerning the flow amplitudes within the interior.
Without better knowledge of the conditions at the convective-radiative 
interface, it is hard to predict precisely what does happen.

While realistic three-dimensional global models of the solar interior 
combining radiative and convective regions are not yet numerically achievable, 
turbulent closure models have been applied to model both azimuthal and 
meridional flows in most of the solar interior 
(Kitchatinov \& Ruediger, 2005; Rempel, 2005). 
Using such a model, Ruediger, Kitchatinov \& Arlt (2005) studied
the penetration of meridional flows into the tachocline. By defining the 
penetration depth $D_{\rm pen}$ as {\it the distance from the base of 
the convection zone to the location of the first reversal in $u_\theta(r)$}, 
they find that $D_{\rm pen} \sim \sqrt{E_\nu}$ and conclude by agreeing 
with GM04 that the depth of penetration of the flows in the interior is indeed
limited to the Ekman solution. However, it is clear from our analysis 
that $D_{\rm pen}$ as defined by Ruediger, Kitchatinov \& Arlt only measures
the variation of the rapidly decaying solution (which indeed must vary 
on an Ekman length). For example, inspection of Figure \ref{fig:rapidsol} 
shows that at a first glance, all of the solutions appear to behave as 
Ekman solutions, while it is only by looking more closely at the flows 
deep in the interior (in Figure \ref{fig:slowsol}) that one can identify 
the presence of the slowly decaying solution with a relatively low but
nonetheless significant amplitude. We suggest that the predictions
of the closure models for flow velocities in the interior could be 
revisited in the light of our analysis. 

Finally, the governing equations in this work have been simplified using 
two major approximations, which must now be briefly addressed. 
Firstly, this study is assumes the flows to be in a quasi-steady state. As
it was shown by McIntyre (2007) this assumption filters out any transient
flows, which could in some circumstances be of much larger amplitude than 
these steady-state flows (see Spiegel \& Zahn, 1992 for example). Therefore
our conclusions are likely to underestimate the actual turnover time of 
transient meridional flows. 

Secondly, 
the momentum and thermal energy equation have been linearized with the 
consequences described in \S\ref{subsec:disc}, an example of which (the 
unphysical counter-rotation) is shown in Figure \ref{fig:nm8}. We do emphasize
that in the real Sun nonlinear effects would play a role in limiting 
the amplitude of the meridional and azimuthal flows penetrating into the 
radiative zone. Therefore our results are consistent with the standard 
theory that no flows with turnover times faster than the thermal diffusion 
time are allowed into the radiative zone. This limits the flow turnover time 
in the tachocline region to a few times $10^5$ years, and into the deep 
interior to a few times $10^7$ years.

Neither of these caveats, however, change the main conclusion of our
analysis, which is that the penetration of meridional flows into
the radiative interior is not necessarily limited to a shallow Ekman depth.

\end{document}